From hoodbhoy@quark.umd.eduFri Feb 13 13:29:09 1998
Date: Tue, 10 Feb 1998 12:06:00 -0500 (EST)
From: Hoodbhoy <hoodbhoy@quark.umd.edu>
To: xdj@quark.umd.edu
Subject: for PRD

\documentstyle[prl,aps,epsfig,preprint]{revtex}
\begin{document}
\tighten

\title{HELICITY-FLIP OFF-FORWARD PARTON DISTRIBUTIONS OF THE NUCLEON}

\author{Pervez Hoodbhoy\footnote{Fulbright scholar, on leave from 
Department of Physics, Quaid-e-Azam University, Islamabad 45320,
Pakistan.} and Xiangdong Ji}
\address{Department of Physics \\
University of Maryland \\
College Park, Maryland 20742 \\
{~}}

\date{U. of MD PP\#98-077~~~DE/ER/40762-142 ~~~ January 1998}

\maketitle
\begin{abstract} 
We identify quark and gluon helicity-flip distributions defined between
nucleon states of unequal momenta. The evolution of these distributions
with change of renormalization scale is calculated in the
leading-logarithmic approximation. The helicity-flip gluon distributions
do not mix with any quark distribution and are thus a unique signature of
gluons in the nucleon. Their contribution to the generalized virtual
Compton process is obtained both in the form of a factorization theorem and
an operator product expansion. In deeply virtual Compton scattering, they can
be probed through distinct angular dependence of the cross section. 
  
\end{abstract}
\pacs{xxxxxx}

\narrowtext

\section{Introduction} 

{}From the point of view of quantum field theory a nucleon is fully
described only if one knows the matrix elements of all possible quark and
gluon operators involving the nucleon state.  Nevertheless, progress is
possible provided one can obtain the matrix element of operators with a
clear physical interpretation. Twist-two operators give the leading
contribution in appropriate hard processes, are relatively simple, and are
more accessible to experimental measurement. The states between which
these operators are sandwiched may be of equal or unequal momenta; the
former situation is familiar from well-investigated processes like deeply
inelastic scattering.  The latter has also been investigated over the
years \cite{ofpd}.  However their importance has been understood only
recently. For instance, knowing certain off-forward twist-two matrix
elements allows for extraction of the quark and gluon orbital and spin
contributions to the nucleon spin \cite{xdj}.  The class of parton
distributions, known as off-forward (off-diagonal, non-forward) parton
distributions defined from these off-forward matrix elements has generated
considerable contemporary interest 
\cite{xdj,rad,xdj1,hood,bal,blu,mel,fran,pet,bel0,bel,man}.  
In simple physical terms, a parton distribution, whether forward or 
off-forward, arises from removal of the parton from the nucleon by a hard 
probe and its subsequent return to form the nucleon ground state further 
along the light-cone, or a similar process. 

The class of twist-two operators which depends on parton 
helicity change shall be the subject of this paper. 
It is now well known \cite{ralston,art,jaf} that there is
one forward chirally odd twist-two proton structure function, 
known as $h_1(x,Q^2)$, measurable in, for example,
the Drell-Yan process. It is defined as the light-cone 
correlation of quark fields weighted by $\sigma^{\mu \nu}$,
\begin{equation}
\int  {d\lambda \over 2\pi} e^{i\lambda x}
      \langle PS'|\bar\psi(-{1\over 2}\lambda n)\sigma^{\mu \nu}
            \psi({1\over 2}\lambda n)|PS \rangle
        = h_1(x) \bar U(PS')\sigma^{\mu \nu} U(PS) + ...\ .  \\
\end{equation} In a helicity basis, wherein spins are measured along the
particle's momentum and $\Sigma_{||}$ is diagonal, $S$ and $S'$ differ by
one unit of angular momentum. Since chirality and helicity coincide for
massless quarks, $h_1$ is a {\em chiral-odd} quantity. To give it a
probability interpretation requires using a {\em transversity} basis
wherein $\Sigma_{\perp}\gamma_0$ is diagonal. This gives $h_1$ the
necessary probabilistic interpretation (for states of equal momenta): it
is the probability to find a quark polarized along the transverse
polarization of the nucleon minus the probability to find the quark
polarized in the opposite direction. There is no gluonic helicity-flip
distribution for obvious reason -- a transverse gluon flipping its
helicity leads to a change of two units of angular momentum and angular
momentum conservation forbids this for a spin-$\frac{1}{2}$ hadron. For
hadrons with spin $\geq 1$ there is no such restriction and some years ago
Jaffe and Manohar \cite{manohar} identified a leading twist gluonic
structure function $\Delta (x,Q^2)$ which can be measured from a
transversely polarized target like the deuteron. 

New Lorentz structures emerge if one allows for off-forward matrix
elements, leading to generalizations of the above mentioned helicity
changing structure functions. Recently, Collins et al. have suggested
measuring the helicity-flip quark distributions in vector meson production
\cite{collins,man1}. The evolution equation for these distributions
has been derived by Belitsky and M\"uller \cite{bel0}. 
Diehl et al. \cite{diehl} have noticed that the
distribution in angle between the lepton and hadron planes in deeply
virtual Compton scattering (DVCS) contains valuable information about the
helicity structure of the nucleon-photon amplitudes.  They point out that
photon helicity flip is possible even with a spin-$\frac{1}{2}$ target
because gluons in the off-forward scattering can transfer two units of
angular momentum. This, of course, requires the existence of the gluon
helicity-flip distributions in the nucleon. Indeed, for off-forward matrix
elements one does not need a state of spin $\geq 1$ to accomodate gluon
helicity flip. 

This paper is intended to present a comprehensive study of leading-twist
helicity-flip off-forward distributions in the nucleon. A systematic
counting suggests that there are four such distributions: two related to
gluon helicity flip and the other two to quark helicity flip.  In each
case, a distribution can be defined depending on whether the nucleon's
helicity is flipped or not. We derive the leading-logarithmic evolution of
these distributions, although in the quark case,
the result was already obtained by Belitsky and M\"uller \cite{bel}. 
In the forward limit, the evolution of the quark
distributions reduces to that of $h_1(x)$ as calculated by Artru and
Mekhfi \cite{art}. The evolution for the gluon distributions reduces to
that of $\Delta(x)$, which has not appeared in the literature before. Note
that the quark helicity-flip distributions do not mix with any gluon ones,
and vice versa.  This is quite significant because, for the first time, we
have a parton distribution that can serve as a unique signature of the
gluons inside the nucleon: gluonic effects cannot be mocked up by any kind
of constituent quarks, and they cannot be evolved away by recklessly
evolving down the momentum scale! 

We study measurement of the helicity-flip gluon distributions in general
two-photon process. The photon helicity-flip Compton amplitude is
calculated in terms of the gluon helicity-flip distributions.  In the
forward limit, we recover the result obtained in \cite{manohar}.  However,
our result is more general. In the language of operator product expansion,
we obtain the leading-order coefficient functions of a class of gluon
operators with total derivatives. According to \cite{diehl}, the
helicity-flip gluon distributions generate distinct angular distributions
in the DVCS cross section. 

The presentation of the paper is as follows. In Section II, we enumerate
the independent helicity amplitudes for the quark-nucleon and
gluon-nucleon sub-processes. Subsequently new helicity changing
distribution functions are motivated and defined.  In Section III the
leading logarithmic evolution of these functions is studied. Section IV
contains a calculation of the Compton amplitude for photon helicity flip
scattering. This amplitude vanishes at the tree level and requires at
least one quark box (plus permutations) to be non-zero. Section V presents
the DVCS cross section that depends on the gluon helicity-flip
distributions. We conclude the paper in Sec. VI. 

\section{ Helicity-Flip Parton Distributions: \\ Counting and Definitions}
   
We shall, in this section, enumerate the complete set of off-forward quark
and gluon distributions at the twist-two level. Helicity-flip ones will
emerge through the counting and will be the focus of this paper.  As
usual, $p^\mu$ and $n^\mu$ are two light-like vectors with $p^2=n^2=0$ and
$p\cdot n=1$. The momenta and spins of the initial and final nucleons are,
respectively, $P,S$ and $P',S'$. The momentum transfer
$\Delta^\mu=P'^\mu-P^\mu$ has both transverse and longitudinal components.
It is convenient to define a special system of coordinates wherein $\bar
P^\mu = (P'+P)^\mu/2$ is collinear and in the $z$ direction,
\begin{eqnarray}
\bar P^\mu &=& p^\mu + (\bar M^2/2) n^\mu  \ , \nonumber \\
\Delta^\mu &=& -2\xi(p^\mu-(\bar M^2/2) n^\mu) 
+ \Delta_\perp^\mu\ ,\nonumber \\
\bar M^2 &=& M^2 -\Delta^2/4 \ .
\end{eqnarray}
The initial nucleon and parton have longitudinal momentum fractions $1+\xi$ 
and $x+\xi$, respectively. 

{}From dimensional reasoning, the leading order contribution to a given hard
process must involve the minimum number of independent parton fields
which, for QCD quantized on the light cone, is two ($\psi_{+}$ for
fermions and $A_\perp$ for gluons). Therefore, one need consider only the
matrix elements of bilinear operators at two different points on the
light-cone. In the kinematic region $x>\xi$, one has the simple
interpretation that the first operator extracts a certain type of parton
from the nucleon and the second replaces it further along the light-cone.
Let $H,H'$ denote the respective helicities of the initial and final
nucleon and $h,h'$ the helicities of the parton extracted and replaced.
The helicity amplitude ${\cal A}_{Hh,H'h'}$ must obey ${\cal
A}_{H'h',Hh}={\cal A}_{Hh,H'h'}$ (time-reversal invariance), and ${\cal
A}_{-H-h,-H'-h'}={\cal A}_{Hh,H'h'}$ (parity invariance). For a purely
collinear process there is no preferred transverse direction; rotational
invariance around the collinear axis requires helicity to be conserved,
$H+h'=H'+h$. However, non-zero transverse momentum of the scattered
nucleon or parton means that, while the total angular momentum will of
course be conserved, helicity conservation will not necessarily hold. The
difference is, of course, absorbed by the orbital motion of the scattered
pair. 

For quarks it is readily seen that a set of independent amplitudes is
provided by the following:  ${\cal A}_{\frac{1}{2} \frac{1}{2},\frac{1}{2}
\frac{1}{2} }$, ${\cal A}_{\frac{1}{2} -\frac{1}{2},\frac{1}{2}
-\frac{1}{2} }$, ${\cal A}_{\frac{1}{2} \frac{1}{2},-\frac{1}{2}
-\frac{1}{2} }$ ${\cal A}_{\frac{1}{2} \frac{1}{2},\frac{1}{2}
-\frac{1}{2} }$, ${\cal A}_{\frac{1}{2} \frac{1}{2},-\frac{1}{2}
\frac{1}{2} }$, and ${\cal A}_{\frac{1}{2} -\frac{1}{2},-\frac{1}{2}
\frac{1}{2} }$.  The familiar distributions $f_1(x,Q^2)$, $g_1(x,Q^2)$,
and $h_1(x,Q^2)$ are linear combinations of the first three in the forward
limit \cite{jaf}.  One unit of orbital angular momentum, made available by
one power of $\Delta_\perp$, allows for three additional amplitudes. A
complete set of off-forward leading-twist quark distributions is given
below: 
\begin{eqnarray}
 \int  {d\lambda \over 2\pi} e^{i\lambda x}
      \langle P'S'|\bar\psi_q(-{\frac{1}{2}\lambda n})\gamma^\mu
            \psi_q(\frac{1}{2}\lambda n)|PS \rangle
        &=& H_q(x,\xi) \bar U(P'S')\gamma^\mu U(PS) \nonumber \\
         && + E_q(x,\xi) \bar U(P'S'){i\sigma^{\mu\nu}
             \Delta_{\nu}
          \over 2M}U(PS) + ...  \ , \nonumber \\
 \int  {d\lambda \over 2\pi} e^{i\lambda x}
      \langle P'S'|\bar\psi_q(-\frac{1}{2}\lambda n)\gamma^\mu\gamma_5
            \psi_q(\frac{1}{2}\lambda n)|PS \rangle
      & =& \tilde H_q(x,\xi)
        \bar U(P'S')\gamma^\mu \gamma_5 U(PS) \nonumber \\
       && + \tilde E_q(x,\xi) \bar U(P'S')
          {\gamma_5\Delta^\mu \over 2M}U(PS) + ...\ , \nonumber \\
 \int  {d\lambda \over 2\pi} e^{i\lambda x}
      \langle P'S'|\bar\psi_q(-\frac{1}{2}\lambda n)\sigma^{\mu \nu}
            \psi_q(\frac{1}{2}\lambda n)|PS \rangle
      & =& H_{Tq}(x,\xi)
        \bar U(P'S')\sigma^{\mu \nu} U(PS) \nonumber \\
       && + E_{Tq}(x,\xi) \bar U(P'S')
          {\gamma^{[\mu}i\Delta^{\nu]} \over M}U(PS) + ...\ , \
\label{quark}
\end{eqnarray}
where $[\mu\nu]$ means antisymmetrization of the two indices and the
ellipses denote higher twist structures which are outside the scope of the
present discussion.  The dependence of each distribution upon $t=\Delta^2$
and $Q^2$ is implicit. In each equation, the first term represents an
amplitude that survives the forward limit and the second term an amplitude
that decouples (but does not vanish) in the forward limit.  The
definitions of $H_q$, $E_q$, $\tilde H_q$, and $\tilde E_q$ are from
Ref.\cite{xdj}. The quark helicity-flip distributions $H_{Tq}$ and
$E_{Tq}$ are new; and they can be selected from the third equation by
taking $\mu=+$ and $\nu=\perp$. The above definitions complete the
identification of all twist-two quark distributions. 

A few additional comments about the definition in Eq. (\ref{quark}) are in
order.  First, for brevity we have not explicitly shown the gauge link
between the quark fields. This link is always present, except in the
light-like gauge $A^+ = 0$. Second, by using time-reversal symmetry, all
the distributions are seen to be real. Third, from taking the complex
conjugate of the above equations, it follows that all the distributions
are even functions of $\xi$. Finally, we can add a time-ordering between
the two fields without changing the content. This follows from,
\begin{eqnarray}
     T\psi_+^\dagger (0) \psi_+(\lambda n) = 
       \psi_+^\dagger(0)\psi_+(\lambda n)
    -\theta(\lambda n^0)\{\psi_+^\dagger(0),\psi_+(\lambda n)\} \ . 
\end{eqnarray}
The second term is just a constant because it is an anticommutator of the
independent (or good) components of the Dirac field separated along the
light cone. Obviously, the constant does not contribute to the matrix
elements. More elaborate but essentially equivalent proofs can be found in
the literature \cite{jaffe,diehl1}. 

We now turn to the gluon distributions. Only transverse gluons need be
considered here because longitudinal ones are either dependent or gauge
degrees of freedom, which lead to either higher twist distributions or
gauge links. An independent set of nucleon gluon amplitudes is:~ 
${\cal A}_{\frac{1}{2} 1,\frac{1}{2} 1 }$, 
${\cal A}_{\frac{1}{2} -1,\frac{1}{2}-1 }$, 
${\cal A}_{\frac{1}{2} 1,-\frac{1}{2} -1 }$ 
${\cal A}_{\frac{1}{2} 1,\frac{1}{2} -1 }$, 
${\cal A}_{\frac{1}{2} 1,-\frac{1}{2} 1 }$, and
${\cal A}_{\frac{1}{2} -1,-\frac{1}{2} 1 }$. 
The familiar distributions $G(x,Q^2)$ and $\Delta G(x,Q^2)$ come from the
forward limit of the first two amplitudes. There is no equivalent of $h_1$
for gluons since it is impossible for a nucleon to spin-flip by two units.
All off-forward twist-two gluon distributions are defined below: 
\begin{eqnarray}
{1\over x} \int { d\lambda\over 2\pi} &&e^{i\lambda x}
  \langle P'S'|F^{(\mu\alpha}(-{\lambda\over2}n)
               F_\alpha^{~\nu)}({\lambda\over2}n)|PS\rangle   
\nonumber \\  && = 
H_g(x,\xi) \bar U(P'S')\bar P^{(\mu} \gamma^{\nu )} U(PS)
 + E_g(x,\xi) \bar U(P'S'){\bar P^{(\mu} i\sigma^{\nu)\alpha }
\Delta_\alpha \over2M} U(PS) + ...\ , \nonumber \\
{1\over x}\int { d\lambda\over 2\pi} && e^{i\lambda x}
  \langle P'S'|F^{(\mu\alpha}(-{\lambda\over2}n)
   i\tilde F_\alpha^{~\nu)}({\lambda\over2}n)|PS\rangle   
\nonumber \\ && =
\tilde H_g(x,\xi) \bar U(P'S')\bar P^{(\mu} \gamma^{\nu)} \gamma_5 U(PS) 
+ \tilde E_g(x,\xi) \bar U(P'S'){\gamma_5\bar 
P^{(\mu}\Delta^{\nu)} \over2M} U(PS)  + ...\ , \nonumber \\ 
{1\over x}\int { d\lambda\over 2\pi}&&e^{i\lambda x}
  \langle P'S'|F^{(\mu\alpha}(-{\lambda\over2}n)
       F^{\nu\beta)}({\lambda\over2}n)|PS\rangle  \nonumber \\ 
 && = 
    H_{Tg}(x,\xi) \bar U(P'S'){\bar P^{([\mu} i\Delta^{\alpha ]}
     \sigma^{\nu\beta)} \over M} 
 U(PS) \nonumber \\  && ~~~~~~~ +~~ 
E_{Tg}(x,\xi) \bar U(P'S'){\bar P^{([\mu} \Delta^{\alpha ]} 
\over M} {\gamma^{[\nu}\Delta^{\beta])} \over M} U(PS) + ... \ . 
\label{gluon}
\end{eqnarray} 
Here in the first two equations, $(\mu\nu)$ means symmetrization of the
two indices and removal of the trace, and in the third equation
$[\mu\alpha]$ and $[\nu\beta]$ are antisymmetric pairs and $(\cdots)$
signifies symmetrization of the two and removal of the trace. These
operations are essential since the product of operators must transform as
irreducible representations of the Lorentz group. 

The distributions $H_g$, $E_g$, $\tilde H_g$, and $\tilde E_g$ have been
introduced before \cite{rad,xdj1}. Their evolutions and mixing with quark
distributions have also been worked out. They play an important roles in
electro-meson production in the small $x$ region \cite{fran,bro}.  The
helicity-flip distributions $E_{Tg}$ and $H_{Tg}$ are new.  They can be
selected from Eq. (\ref{gluon}) by taking $\mu=\nu=+$ and $\alpha,\beta =
\perp$. The angular momentum conservation requires presence of one unit of
angular momentum ($\Delta_T$) when the nucleon helicity is flipped
($H_{Tg}$) and two units ($\Delta_T\Delta_T$) when it is not ($E_{Tg}$).
Hence both decouple from the matrix elements in the forward limit although
the distributions themselves do not vanish. 

In a sense, the gluon helicity-flip distributions are the ``cleanest"
among the class of gluon distributions since they are forbidden to mix
with quark distributions by angular momentum conservation. This, in fact,
was why Jaffe and Manohar \cite{manohar} had proposed using the
$\Delta(x,Q^2)$ distribution as a probe of ``exotic gluons" in a spin $J\ge
1$ nucleus.  Nuclear binding or pions would not contribute. However there,
unlike here, there is a strong suppression on the magnitude of the
distribution due to the small size of the nuclear interaction relative to
a typical hadronic mass scale because ``exotic gluons" can only be
generated by the nuclear interaction (for an estimate of $\Delta(x,Q^2)$
see Ref.(\cite{nzar}) ). In the nucleon, the off-forward helicity-flip
gluon distributions can be as large as other gluon distributions for a
reasonable size of $x$. It is interesting to note that the n=2 moment of 
the gluon distributions above can be directly expressed in terms of the 
matrix elements of the colour electric and magnetic fields in the nucleon.

\section{ Evolution Of Helicity-Flip Distributions}

A parton distribution is necessarily defined at a given distance or
momentum scale because of ultraviolet divergences. Physically, the scale
can be related to the kinematic variables of the particular hard process
under consideration. In this section we study, using momentum space
Feynman diagrams, the leading-logarithmic evolution of the helicity-flip
parton distributions defined in the previous section. While the method is
straightforward in principle, there are important subtleties related to
the gauge dependence of the calculation and end-point singularities. In
principle any gauge choice is permissible but the light-cone gauge is
naturally preferred for several reasons: the fewest number of diagrams
need be calculated, path-ordered exponentials are absent, the light-cone
dynamics of partons is transparent, and ghosts are absent. However, there
is a price to be paid because the light-cone gluon propagator, when
imbedded in a loop, leads to singularities in the end-points of integrals
whose interpretation is ambiguous. This is sufficient reason to do the
(longer) calculation in a covariant gauge. Therefore, in the following, we
shall use the Feynman gauge and treat separately the quark and gluon
helicity-flip distributions. At leading-logarithmic order, any ultraviolet
regulator is as good as any other; so we impose an ultraviolet cut-off on
the momentum integrations.

\subsection{Evolution of the Quark Distributions $E_{T}$ and $H_{T}$}

The result in this section has been obtained before in Ref. \cite{bel0}.
Here for completeness we present our calculation in a different form.

{}From the definitions of the quark distributions $H_{Tq}$ and $E_{Tq}$ in
Eq.(\ref{quark}), it is clear that the two will evolve identically with
the leading component of the operator $\bar \psi \sigma^{\mu \nu} \psi$. 
Selecting only the leading twist part, it is therefore convenient to
define,
\begin{equation}
{\cal F}(x,\xi,Q^2)=n_\mu e_\nu \int  {d\lambda \over 2\pi} e^{i\lambda x}
      \langle P'S'|\bar\psi(-{\frac{1}{2}\lambda n})\sigma^{\mu \nu}
e^{-ig\int^{-\lambda/2}_{\lambda/2} d\alpha n\cdot A(\alpha n)}
            \psi(\frac{1}{2}\lambda n)|PS \rangle .
\label{Kdef}
\end{equation}
where $e^\mu$ is a unit vector in a transverse direction.  The
path-ordered integral has been reinstated in the above.  The
renormalization scale $Q^2$ is the cut-off for the momentum components of
the fields.  ${\cal F}(x,\xi,Q^2)$ can be diagrammatically represented as
in Fig.1a. Now imagine a slight increase of $Q^2$ to $Q^2 +\delta Q^2$,
revealing a deeper level of hadronic substructure.  Additional diagrams
contributing to ${\cal F}(x,\xi,Q^2+\delta Q^2)$ are shown in Figs. 1b-1e. 

\begin{figure}
\label{fig1}
\epsfig{figure=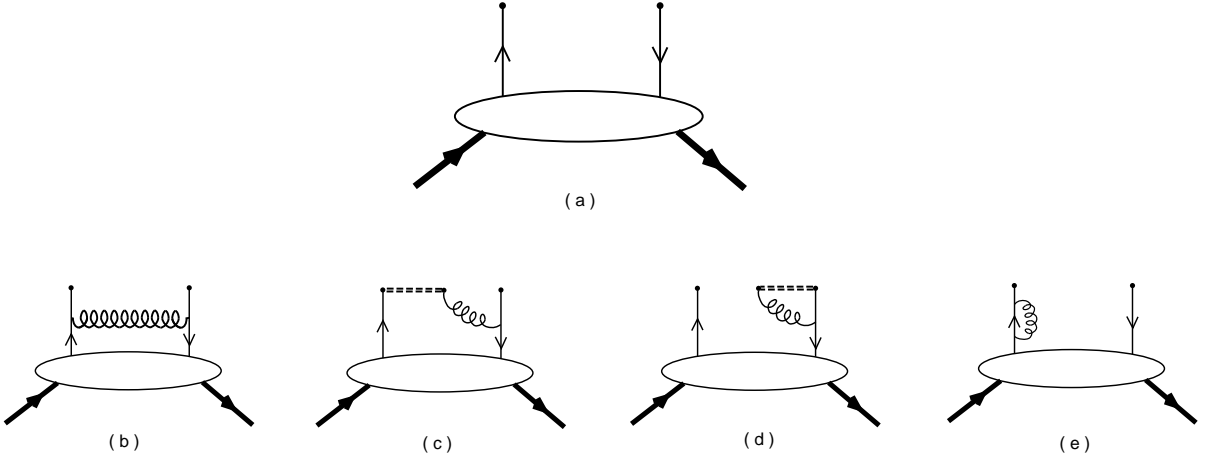,height=6.0cm}
\caption{Feynman diagrams for evolution of quark
helicity-flip parton distributions.}
\end{figure}
 
Feynman-like rules for these diagrams can be derived in a rather
straightforward way: the product of bare Heisenberg fields in
Eq.(\ref{Kdef}) can be brought under the time-ordering symbol and a
perturbation expansion follows from expanding out the exponential
containing the QCD interaction terms $\exp(i\int{\cal L}_{\rm int})$ as
well as expanding the path-ordered exponential. Subsequently a collinear
expansion is made:  the lines entering or leaving the hadron blob are
always collinear, reflecting the fact that the internal relative momenta
of partons in the hadron are much less than the momentum of the hard
probe. The transverse momenta of the other lines is bounded by $Q^2$. 

The Feynman expressions for fig.1c and 1d are,
\begin{eqnarray}
\delta {\cal F}_c(x,\xi,Q^2) &=& \frac{1}{2}n_{[\mu}e_{\nu ]}\int dy 
\int{d^4k\over (2\pi)^4} 
\delta (x-k\cdot n) {i\over (q-k)\cdot n +i\epsilon} iD_{\alpha\beta}(q-k)
\nonumber \\
&& \times {\rm Tr}\left[ \frac{1}{8}\sigma^{\mu\nu}(-igt^an^\alpha)
\sigma^{\rho\lambda}(-igt^a\gamma^\beta)iS_F(k+\frac{1}{2}\Delta)
\right ] p_{[\rho} e_{\lambda]}{\cal F}(y,\xi,Q^2) \ ,\\
\delta {\cal F}_d(x,\xi,Q^2) &=& \frac{1}{2}n_{[\mu}e_{\nu ]}\int dy
\int {d^4k\over (2\pi)^4}
\delta (x-y) {i\over (k-q)\cdot n +i\epsilon} iD_{\alpha\beta}(q-k)
\nonumber \\
&& \times {\rm Tr}\left[ \frac{1}{8}\sigma^{\mu\nu}(-igt^an^\alpha)
\sigma^{\rho\lambda}(-igt^a\gamma^\beta)iS_F(k+\frac{1}{2}\Delta)
\right ] p_{[\rho} e_{\lambda]}{\cal F}(y,\xi,Q^2) \ .
\end{eqnarray}
In the first of the above two equations, $q^\mu=yp^\mu$ and in the second,
$q^\mu=xp^\mu$. Two other diagrams are obtained by reflection and yield
identical expressions under $\xi\rightarrow -\xi$. Finally, the quark
self-energy diagram in fig.1e (and its reflection), together with the
wavefunction renormalization, yield the combined expression below after 
calculation,
\begin{eqnarray}
{ D_Q {\cal F}(x,\xi,Q^2)\over D\ln Q^2} &=& 
{\alpha_s(Q^2) \over 2\pi}C_F \left[   
\theta (x-\xi)\int^1_x  {dy\over y} {x-\xi\over y-\xi} +
\theta (x+\xi)\int^1_x  {dy\over y} {x+\xi\over y+\xi}~~  \right. \nonumber \\ 
&& \left. - \theta (\xi-x)\int^x_{-1}  {dy\over y} {x-\xi\over y-\xi} -
\theta (-\xi-x)\int^x_{-1}  {dy\over y} {x+\xi\over y+\xi} 
\right ]{{\cal F}(y,\xi,Q^2)\over 
y-x+i\epsilon} \ , 
\label{apq}
\end{eqnarray}
where,
\begin{equation}
     {D_Q\over D\ln Q^2} = {d\over d\ln Q^2}
       - {\alpha_s(Q^2)\over 2\pi}C_F
      \left[{3\over 2} + \int^x_{\xi} {dy\over
        y-x-i\epsilon} + \int^x_{-\xi}{dy \over
        y-x-i\epsilon}\right] \ .
\end{equation}
The above result agrees with that obtained by Belitsky and
M\"uller. 

It is useful to take moments of ${\cal F}(x,\xi,Q^2)$. Define,
\begin{equation}
{\cal F}_n (\xi,Q^2)=\int^1_{-1} dx~x^{n-1}{\cal F}(x,\xi,Q^2)~~~~(n\ge 2)\ , 
\end{equation}
where $n=$even (odd) moments are charge conjugation even (odd). Then, a
calculation leads to the following evolution equation for the moments,
\begin{equation}
{ d {\cal F}_n\over d\ln Q^2}= {\alpha_s(Q^2) \over 2\pi}C_F \left[
\left(\frac{3}{2}-2S(n)\right){\cal F}_n + 
2\sum^{[\frac{n-1}{2}]}_{i=1,2,..}(\frac{1}{2i}-\frac{1}{n})(2\xi)^{2i}
{\cal F}_{n-2i} \right] \ , 
\label{mom}
\end{equation}
where,
\begin{equation}
S(n)=\sum^n_{i=1}\frac{1}{i} \ . 
\end{equation}
We shall now interpret the evolution equation in terms of operator 
mixing. To this end, define,
\begin{eqnarray}
{\cal O}^{\mu_1 \cdots \mu_n\nu}_{n,2i}(x) = i\partial^{\mu_1}\cdots 
i\partial^{\mu_{2i}} \bar\psi \stackrel{\leftrightarrow}
{iD}^{\mu_{2i+1}} \cdots \stackrel{\leftrightarrow}{iD}^{\mu_{n-1}} 
\sigma^{\mu_n\nu}\psi \ . 
\label{op2}
\end{eqnarray}
where all $\mu$ indices are symmetrized and
$\stackrel{\leftrightarrow}{iD} = [\stackrel{\rightarrow}{iD}
 -\stackrel{\leftarrow}{iD}]/2$. 
Using translational invariance, it is easy to see that,
\begin{equation}
n_{\mu_1}\cdots n_{\mu_{n}}e_\nu \langle P'S'|{\cal O}^{\mu_1 \cdots 
\mu_n\nu}_{n,2i} |PS\rangle = (2\xi)^{2i}{\cal F}_{n-2i} \ .
\end{equation}
As one can see from Eq.(\ref{mom}), the operators ${\cal O}^{\mu_1 \cdots
\mu_n\nu}_{n,2i}$ belonging to same $n$ but different $i$ mix.  Taking an
appropriate linear combination of these, or equivalently, diagonalizing
the mixing matrix, is not difficult. One may establish recursion relations
using Eq.(\ref{mom}) to finally arrive at,
\begin{eqnarray} 
{ d {\tilde{\cal O}}_n \over d\ln Q^2} &=& {\alpha_s(Q^2) 
\over 2\pi}C_F \left( \frac{3}{2}-2S(n)\right) {\tilde{\cal O}}_n \ , \\ 
{\tilde{\cal O}}_n &=&\sum^{[\frac{n-1}{2}]}_{i=0} {(-1)^i~2^{n-2i-1}~ 
\Gamma(n-i+\frac{1}{2}) \over (n-2i-1)!~i!~\Gamma(\frac{3}{2})}
{\cal O}_{n,2i}^{\mu_1\cdots\mu_n\nu}  \ .  
\label{geg} 
\end{eqnarray}
The coefficients in Eq.(\ref{geg}) are those of the Gegenbauer
polynomials, $C^{\frac{3}{2}}_{n-1}(x)$. This can be traced to the fact
that these polynomials are essentially the Clebsch-Gordan coefficients
which occur in the light-cone expansion of operator products that
transform irreducibly under the conformal group \cite{gatto}. 

\subsection{Evolution of the Gluon Distributions $E_{T}$ and $H_{T}$}

The evolution of the gluon helicity-flip distributions can be studied in
the same way as for the quark.  For convenience, we define,
\begin{eqnarray}
&& {\rm F}(x,\xi,Q^2)=n_\mu n_\nu e_{(\alpha} e'_{\beta)} \nonumber \\
&& \times \int  {d\lambda \over 2\pi} e^{i\lambda x}
      \langle P'S'|F^{\mu\alpha}(-{{1\over 2}\lambda n})
e^{-ig\int^{-\lambda/2}_{\lambda/2} d\alpha n\cdot A(\alpha n)}
            F^{\nu\beta}(\frac{1}{2}\lambda n)|PS \rangle .      
\end{eqnarray}
where $e_\alpha$ and $e_\beta'$ are two unit vectors in the transverse
directions. We are interested in the change of $\rm F$ under the change of
the momentum cut-off $Q^2$.  Some representative Feynman diagrams are
shown in Fig.2. 

\begin{figure}
\label{fig2}
\epsfig{figure=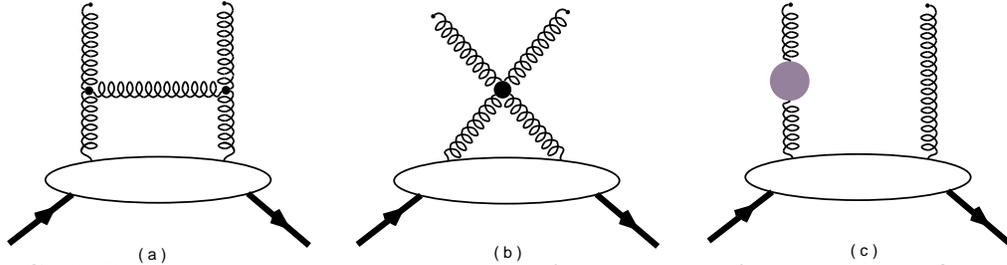,height=3.5cm}
\caption{Representative Feynman diagrams for evolution of gluon
helicity-flip parton distributions.}
\end{figure}  

As we have mentioned before, the gluon helicity-flip distributions do not
mix with any quark distributions. For $x>\xi$, the evolution equation
reads,
\begin{equation}
     { D_Q {\rm F}(x,\xi,Q^2) \over D \ln Q^2}
   = {\alpha_s(Q^2)\over 2\pi}
   \int^1_x {dy\over y} P\left({x\over y}, 
    {\xi\over y}, {\epsilon\over y}\right) {\rm F}(y,\xi,Q^2) \ , 
\end{equation}
where, 
\begin{equation}
     {D_Q\over D\ln Q^2} = {d\over d\ln Q^2}
       - {\alpha_s(Q^2)\over 2\pi}C_A
      \left[{11\over 6} - {n_f\over 3C_A} + \int^x_{\xi} {dy\over
        y-x-i\epsilon} + \int^x_{-\xi}{dy \over
        y-x-i\epsilon}\right] \ ,
\end{equation}      
and,
\begin{equation}
       P(x,y,\epsilon) = 2C_A {(x^2-\xi^2)\over 
       x(1-\xi^2)^2}\left({2(1-x)\xi^2\over x^2-\xi^2}
       +{1-\xi^2\over 1-x+i\epsilon} \right) \ . 
\end{equation}
For $-\xi<x<\xi$, the equation is,
\begin{equation}
     { D_Q {\rm F}(x,\xi,Q^2) \over D \ln Q^2}
   = {\alpha_s(Q^2)\over 2\pi}\left(
   \int^1_x {dy\over y} P'\left({x\over y},
    {\xi\over y}, {\epsilon\over y}\right) 
   - \int^x_{-1} {dy\over y}
    P'\left({x\over y},
    - {\xi\over y}, {\epsilon\over y}\right)\right) 
  {\rm F}(y,\xi,Q^2) \ , 
\end{equation}    
where,
\begin{equation}
     P'(x,\xi,\epsilon)
   = {(x^2-\xi^2) \over x (1-\xi^2) }
    \left[ {2\xi\over (x-\xi)(1+\xi)} 
   + {4\over 1-x+i\epsilon} \right] \ . 
\end{equation}
And finally for $x<-\xi$, the equation is the same as that for
$x>\xi$, except $\int^1_x \rightarrow -\int^x_{-1}$. 

In the forward limit, the evolution equation reduces to
that for $\Delta(x)$, 
\begin{equation}
      {d\Delta(x)\over d\ln Q^2} = 
    {\alpha_s(Q^2)\over 2\pi}\int^1_x {dy\over y}
    P\left({x\over y}\right) \Delta(y)\ , 
\end{equation}
where the evolution kernel is,
\begin{equation}
        P(x) = 2C_A {x\over (1-x)_+} + \left({11C_A\over 6}
       - {n_f\over 3}\right)\delta(x-1) \ .  
\end{equation}
Here the + prescription is standard \cite{ap}.  The above result, as far
as we know, is new. 

It is instructive to look at the evolution in operator form. 
Define the $n=$ even moments (the odd moments vanish
because ${\rm F}(x)$ is antisymmetric in $x$
as one can easily check from the definition),
\begin{equation}
    {\rm F}_n(\xi, Q^2) = 
   \int^1_{-1} dx x^{n-1} {\rm F} (x,\xi, Q^2) ~~~~(n\ge 2) \ . 
\end{equation}
The evolution equation becomes,
\begin{equation}
      {d {\rm F}_n\over d \ln Q^2}
    = {\alpha_s(Q^2)\over 2\pi} C_A\left[
       \left({11\over 6} - {n_f\over 2C_A} -2S(n)\right){\rm F}_n
         + \sum^{[{n-1\over 2}]}_{i=1} \left({4i-2\over n-1}
      - {4i+2\over n} + {1\over i}\right)(2\xi)^{2i} {\rm F}_{n-2i}
        \right] \ .
\end{equation}
Define a tower of twist-two gluon operators,
\begin{equation}
     {\cal O}^{\mu_1\cdots\mu_n\alpha\beta}_{n,2i} 
     = i\partial^{\mu_1}\cdots 
      i\partial^{\mu_{2i}}F^{\mu_{2i+1}\alpha}\stackrel{\leftrightarrow}
      {iD}^{\mu_{2i+2}}
    \cdots \stackrel{\leftrightarrow}{iD}^{\mu_{n-1}}F^{\mu_n \beta} \ . 
\label{gluop}
\end{equation}
Then it is easy to see,
\begin{equation}
       n_{\mu_1}\cdots n_{\mu_n} e_\alpha e_\beta'
      \langle P'S'|{\cal O}^{\mu_1\cdots\mu_n\alpha\beta}_{n,2i}|PS\rangle
    = (2\xi)^{2i} {\rm F}_{n-2i} \ . 
\end{equation}
Hence the mixing of the different moments of the gluon distributions
reflects the mixing of the twist-two operators of same spin and dimension.
Define a new basis of operators in term of the Gegenbauer polynomials
$C_{n-2}^{5\over 2}(x)$ combinations,
\begin{equation}
{\cal O}_n = \sum^{[\frac{n-2}{2}]}_{i=0} {(-1)^i~2^{n-2i-2}~
\Gamma(n-i+\frac{1}{2}) \over (n-2i-2)!~i!~\Gamma(\frac{5}{2})}
{\cal O}_{n,2i}^{\mu_1\cdots\mu_n\alpha\beta} \ . 
\end{equation}
Then the evolution of ${\cal O}_n$ becomes diagonal,  
\begin{equation}
           { d{\cal O}_n  \over d \ln Q^2}
        = {\alpha_s(Q^2)\over 2\pi} C_A\left({11\over 6}
         - {n_f\over 3C_F} - 2S(n)\right) {\cal
          O}_n \ . 
\end{equation}
Again, the above simplification is due to conformal symmetry. However,
beyond the leading-logarithmic order, the conformal symmetry is
anomalously broken by quantum corrections \cite{mul}. 

\section{photon Helicity-flip Compton amplitude}

The gluon helicity-flip distributions, $H_{Tg}(x,\xi)$ and
$E_{Tg}(x,\xi)$, are basic properties of the nucleon, at par with the
other distributions, namely, $H_g(x,\xi)$, $E_g(x,\xi)$, $\tilde
H_g(x,\xi)$, and $\tilde E_g(x,\xi)$.  Because of angular momentum
conservation, the gluon double-helicity flip distributions do not mix with
quark distributions, making their isolation and possible measurement
relatively cleaner. One would like to know which hard processes probe
$H_{Tg}$ and $E_{Tg}$. The general Compton scattering involving two
photons offers one possibility, perhaps the simplest.  Diffractive vector
meson production from a deeply-virtual photon is another
\cite{collins,bro}. 

\begin{figure}
\label{fig3}
\epsfig{figure=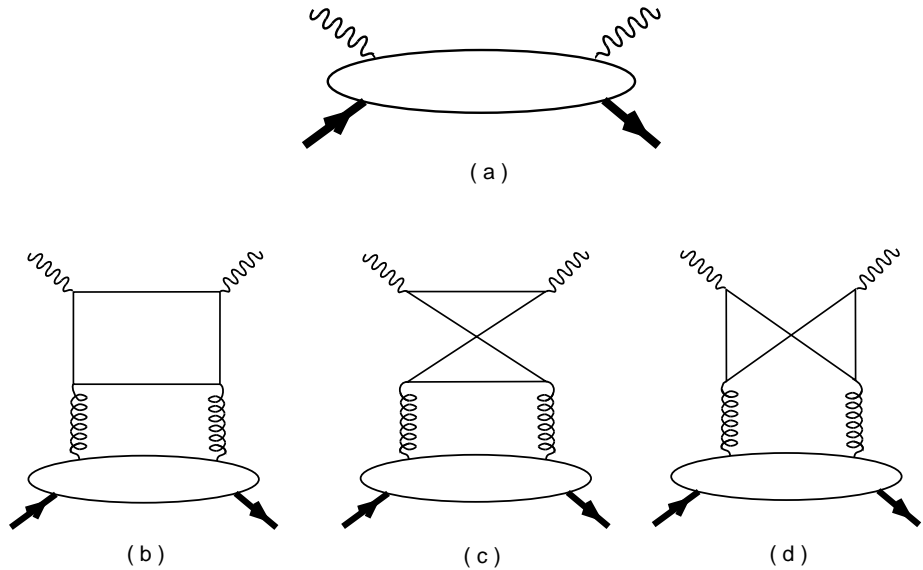,height=7.5cm}
\caption{Diagrams for photon helicity-flip Compton scattering.}
\end{figure}   

Fig. 3a illustrates the general two photon process.  A highly virtual
photon with momentum $q+\Delta/2$ and transverse polarization is incident
upon the nucleon. A second photon of momentum $q-\Delta/2$ is detected,
and the recoiling nucleon emerges intact. The scattering amplitude is,
\begin{equation}
T^{\mu\nu}=i\int d^4z~e^{-iq\cdot z}
\langle P'S'|T\left[ J^\nu(-\frac{1}{2}z)J^\mu(\frac{1}{2}z)\right] 
|PS \rangle \ . 
\end{equation}
A convenient set of kinematic variables is obtained by choosing $q^\mu$
and $\bar P^\mu = \frac{1}{2}(P+P')^\mu$ to be collinear and in the 3
direction. In terms of $p^\mu$, $n^\mu$, we expand the ``average" photon
momentum,
\begin{eqnarray}
          q^\mu = -x_B p^\mu + (Q^2/2x_B) n^\mu  \ .  
\end{eqnarray}
The transverse components of $\Delta^\mu$ are assumed to be of the order
of the nucleon mass, i.e. much smaller than the hard momentum $Q$.  In the
collinear approximation, the momenta of the incoming and outgoing photons
are, respectively, $q-\xi p$ and $q+\xi p$.  The initial and final state
nucleon's momenta are $(1+\xi)p^\mu$ and $(1-\xi)p^\mu$, respectively. 
Symmetrizing the diagrams in this manner makes it possible to get simpler
expressions by exploiting crossing symmetry. 

There is no tree-level coupling between the scattering photon and the
gluons in the target. However, there is a contribution at $O(\alpha_s)$
from the box-diagram and its permutations, shown in Figs. 3b-d. The
momenta of the gluons going outward and returning to the nucleon blob are,
respectively, $(x+\xi)p^\mu$ and $(x-\xi)p^\mu$ with $-1<x<1~$. The loop
momentum is expanded as,
\begin{equation}
 k^\mu = (k\cdot n)p^\mu + (k\cdot p )n^\mu + k^\mu_\perp \ \nonumber~.
\end{equation}
Rather than use Feynman parameters to solve the integral with four
products in the denominator, we systematically exploit crossing symmetry
$(\xi \rightarrow -\xi)$ to reduce the integral to a much more manageable
form with three product terms.  Dimensional regularization is used for the
divergent graphs. The infrared singularity cancels in the sum of the three
diagrams, as it must. Many details of the calculation are similar to those
in Ref.\cite{osborne}. 
 
Our final result for the photon double helicity-flip amplitude is, 
\begin{equation} 
\label{prod} 
T^{\mu\nu} = {\alpha_s\over 2\pi} \left(\sum_q e_q^2\right)
      \int^1_{-1} dx {x\over x^2- \xi^2}
\left[1+
{x_B^2-\xi^2\over x^2-\xi^2} \ln \left({x^2_B-x^2 \over 
x_B^2-\xi^2}\right)~\right] n_\alpha n_\beta
{\cal T}^{\mu\nu\alpha\beta} \ , 
\end{equation}
where,
\begin{equation}
    {\cal T}^{\mu\nu\alpha\beta}
    = {1\over x}\int { d\lambda\over 2\pi}e^{i\lambda x}
  \langle P'S'|F^{(\mu\alpha}(-{\lambda\over2}n)
       F^{\nu\beta)}({\lambda\over2}n)|PS\rangle \ .  
\end{equation}
In the above, $q$ sums over the quarks circulating in the loop and $e_q$
are their electric charges. Eq.(\ref{prod}) is the convolution of a
perturbatively calculable part and a soft part reflecting the nucleon's
composition, and is in the form of a factorization theorem at the lowest
order of QCD perturbation theory. To completely justify the use of the
soft part, one needs to consider all contributing Feynman diagrams with
the hard quark loop connecting with arbitrary number of gluons to the
nucleon blob. Any choice of gauge may be made of course, but the final
result will be gauge independent. It is easiest to work in the light-cone
gauge and with simply two physical gluon fields. In covariant gauge it
would be necessary to show that these additional gluons are summed up in
the path-ordered integral. 

As a check on our calculation, we may consider forward scattering $\xi=0$
on a target with spin $J\ge 1$. In the limit of $x_B\rightarrow \infty$,
the perturbative part in Eq.(\ref{prod}) can be expanded: 
\begin{eqnarray}
T^{\mu\nu} &=& -{\alpha_s\over 2\pi}~\left(\sum_q e_q^2\right)~
       \sum_{n ~{\rm even}=2}^\infty ~{1\over x_B^n} ~
      {2\over n+2}~ \int^1_{-1}dx~ x^{n-1}
       {\cal T}^{\mu\nu\alpha\beta}n_\alpha n_\beta \nonumber \\
      & = & -{\alpha_s\over 2\pi} ~\left(\sum_q e_q^2\right)~
       \sum_{n=2}^\infty ~
      {2\over n+2} ~{2^n q^\mu_1 \cdots q^\mu_n \over (Q^2)^n}~
       \nonumber \\ && \times 
       \langle PS|F^{\mu\mu_1}\stackrel{\leftrightarrow}
      {iD}^{\mu_2} \cdots \stackrel{\leftrightarrow}
      {iD}^{\mu_{n-1}} F^{\nu\mu_n} |PS\rangle
\label{op1}
\end{eqnarray}
The above coincides with the result in Ref.\cite{manohar} except for an
overall sign. In the general case, we can convert Eq.(\ref{prod}) result
into a generalized operator production expansion with derivative
operators, 
\begin{eqnarray}
    i\int d^4z e^{-iqz} J^\nu(-{z\over 2})
     J^\mu({z\over 2}) &= &  -{\alpha_s\over 2\pi} 
     ~\left(\sum_q e_q^2\right)~
       \sum_{n~{\rm even}=2}^\infty ~
      {2\over n(n+2)} ~{2^n q^\mu_1 \cdots q^\mu_n \over (Q^2)^n} 
      \nonumber \\ && \times
    \sum_{i=0}^{\left[{n-1\over 2}\right]}
     (n-2i) {\cal O}_{\mu_1\cdots\mu_n\mu\nu}^{n,2i} + ...    
\end{eqnarray}
where ${\cal O}_{\mu_1\cdots\mu_n\mu\nu}^{n,2i}$ is defined in
Eq.(\ref{gluop}). We would like to emphasize again that only the $\mu\nu$
symmetric and traceless terms are included in the above equation.

\section{ DVCS cross section with gluon helicity-flip distributions}

The Compton amplitude in the last section can be used to obtain the cross
section for deeply virtual Compton scattering. In DVCS, the final photon
is real and hence one has the constaint $x_B=\xi$. In this section, we
calculate the cross section in this special kinematic limit. 

We choose the kinematic variables as those used in Ref.\cite{xdj1}:
$k=(\omega, \vec{k})$ and $k'=(\omega', \vec{k'})$ for the four-momenta of
the intial and final electrons, $P = (M,0)$ and $P'=(E',\vec{P'})$ for the
initial and final momenta of the nucleon, and $q'=(\nu', \vec{q'})$ the
momentum of the final photon. The differential cross section is,
\begin{equation}
     d\sigma = |{\cal T}|^2 d\Gamma \ , 
\end{equation}
where $\cal T$ is the invariant $T$ matrix of the scattering and $d\Gamma$
is the invariant phase space factor. Depending on choice of independent
kinematic variables to characterize the differential cross section,
$d\Gamma$ takes different form. If one uses the scattered electron's
energy and solid angle, and the scattered nucleon's solid angle,
\begin{equation}
    d\Gamma = {1\over 32(2\pi)^5 \omega M}
           \omega' d\omega' d\Omega_{e'} d\Omega_{P'}
            {  P'^2\over |P'(\nu+M) - qE'\cos\phi|} \ ,
\end{equation} 
where $\phi$ is the angle between $\vec{q}$ and $\vec{P'}$, and the sum
over two possible solutions of $|\vec{P'}|$ is implicit. On the other
hand, one can also use the standard $Q^2$ and $x_B$ (or $s$),
$t=\Delta^2$, the $t$-channel momentum transfer, and $\phi$ the angle
between lepton and hadron planes \cite{diehl,kro}. 

The invariant $T$ matrix consists of two parts. The first part comes from
the Compton scattering,
\begin{equation}
     {\cal T}_1  = - e^3\bar u(k')\gamma^\mu u(k)
    {1\over q^2} T_{\mu\nu} \epsilon^{\nu *}\ ,
\end{equation}
where $\bar u, u$ are the spinors of the lepton and $\epsilon$ is the
polarization of the emitting photon. The Compton amplitude $T^{\mu\nu}$
contains both photon helicity-flip and non-flip contributions,
\begin{equation}
     T_{\mu\nu} = T_{\mu\nu}^{\Delta \lambda =0}
            + T_{\mu\nu}^{\Delta \lambda = 2} \ , 
\end{equation}
where the first term is given by Eq.(4) in Ref.\cite{xdj1} and the second 
term is from Eq.(\ref{prod}), \begin{eqnarray}
      T^{\mu\nu}_{\Delta \lambda = 2}    
     &=& {\alpha_s\over 4\pi} ~\left(\sum_i e_i^2\right)
      \int^1_{-1}dx\left({1\over x-\xi+i\epsilon}
      + {1\over x+\xi - i\epsilon}\right)
      \nonumber \\ &&
       \times ~n_\alpha n_\beta \left[
       H_{Tg}(x,\xi) \bar U(P'S'){\bar P^{([\mu} i\Delta^{\alpha ]}_\perp
     \sigma^{\nu\beta)} \over M} U(PS) \right. \nonumber 
    \\  &&  \left.  ~~~~~~~~~~~+ ~E_{Tg}(x,\xi) \bar 
   U(P'S'){\bar P^{([\mu} \Delta^{\alpha ]}_\perp
   \over M} {\gamma^{[\nu}\Delta^{\beta])}_\perp \over M} U(PS)\right]
   \nonumber \\  
    &=& {\alpha_s\over 4\pi} ~\left(\sum_q e_q^2\right)
      \int^1_{-1}dx\left({1\over x-\xi+i\epsilon}
      + {1\over x+\xi - i\epsilon}\right) {1\over 4M}
      \nonumber \\ &&
       \times \left[ 
        H_{Tg}(x, \xi)\bar U(P')(\Delta^\mu_\perp \gamma^\nu_\perp
       + \Delta^\nu_\perp \gamma^\mu_\perp - g^{\mu\nu}_\perp 
	\not\!\Delta_\perp)\not\! n U(P) \right. \nonumber \\ 
       && \left. +{1\over M} E_{Tg}(x,\xi)
     \bar U(P')\left(\xi(\Delta^\mu_\perp\gamma^\nu_\perp
	+\Delta^\nu_\perp\gamma^\mu_\perp
               - g^{\mu\nu}_\perp \not\! \Delta_\perp)
        + \not\! n (\Delta^\mu_\perp \Delta^\nu_\perp - {1\over 2}g^{\mu\nu}
            \Delta_\perp^2) \right)U(P)\right] \ . 
\label{long}
\end{eqnarray}
The second part of the T matrix comes from the Bethe-Heitler process,
\begin{equation}
       {\cal T}_2 = -e^3 \bar u(k')\left[\not\! \epsilon^*{1\over
      \not\! k -\not\!\Delta-m_e +i\epsilon}\gamma^\mu +
     \gamma^\mu {1\over \not\! k' + \not\! \Delta - m_e +i\epsilon}
     \not\!\epsilon^*\right]u(k) {1\over \Delta^2} \langle P'
    |J_\mu(0)|P\rangle \ ,
\end{equation}
where $m_e$ is the mass of electron and will be ignored for the following
discussion. The elastic nucleon matrix element is, 
\begin{equation}
   \langle P'|J_\mu(0)|P\rangle = \bar U(P')\left[\gamma_\mu 
    F_1(\Delta^2) + F_2(\Delta^2) {i\sigma_{\mu\nu} 
  \Delta^\nu \over 2M} \right] U(P) \ ,
\end{equation}
where $\bar U, U$ are the nucleon spinors and $F_1$ and $F_2$ are the
usual Dirac and Pauli form factors of the nucleon. 
                           
We are interested in only the leading contribution to the cross section
from the helicity-flip gluon distributions. This comes from the
interferences between the helicity-flip and non-flip Compton amplitudes
and between the former and the Bethe-Heitler amplitude.  The first
interference yields,
\begin{equation}
       \left({\cal T}_1^{\Delta \lambda=0 }\right)^*
      {\cal T}_1^{\Delta \lambda=2} + 
        {\cal T}_1^{\Delta \lambda=0}
      \left({\cal T}_1^{\Delta \lambda=2}\right)^* 
    =  - {e^6\over Q^4} \ell^{\mu\nu}_{\rm VC}
   W_{\rm VC\mu\nu} \ ,
\end{equation} 
where the lepton tensor $\ell^{\mu\nu}_{\rm VC}$ can be found in Ref.
\cite{xdj1}. The hadron tensor is
\begin{eqnarray}
    W_{\rm VC}^{\mu\nu} & = & {1\over M^2} 
     (\Delta^\mu_\perp \Delta^\nu_\perp - 
      {1\over 2}g^{\mu\nu}_\perp \Delta_\perp^2)
     {\alpha_s\over 4\pi}\left(\sum_q e_q^2\right) \nonumber \\ &&
    \times \sum_q e_q^2 ~{\rm Re}\int^1_{-1} dx \alpha(x) \int^1_{-1} dx' \alpha^*(x')
     \Big[H_q(x,\xi)E_{Tg}(x',\xi)-H_{Tg}(x',\xi)E_q(x,\xi)
     \Big] \ . 
\end{eqnarray}
The tensor structure $(\Delta^\mu_\perp \Delta^\nu_\perp - {1\over
2}g^{\mu\nu} \Delta_\perp^2)$ signals a $\cos 2\phi$ term in the cross
section, as was noted in \cite{diehl}. There is, 
of course, also a gluon helicity non-flip 
term but it enters at one higher power of $\alpha_s$.

The interference between the double-helicity-flip Compton and
Bethe-Heitler amplitudes is,
\begin{equation}
         {{\cal T}_1^{\Delta \lambda =2}}^*{\cal T}_2 
          + {\cal T}_1^{\Delta \lambda =2}{\cal T}_2^*
     = 2{e^6\over \Delta^2 Q^2}~\ell^{(\mu\nu)
         \alpha}~{\rm Re} H_{(\mu\nu)\alpha} \ , 
\end{equation}               
where $\ell^{(\mu\nu)\alpha}$ depends on electron kinematic variables and
can be found in \cite{xdj1}. The nucleon
structure dependent part is,
\begin{eqnarray}
   H^{(\mu\nu)\alpha}
  &=& \left(\Delta^\mu_\perp \Delta^\nu_\perp - {1\over 2} 
    g^{\mu\nu}_\perp \Delta^2_\perp\right) 
    { \alpha_s\over 4\pi} \left(\sum_q e_q^2\right) \int^1_{-1} dx \alpha(x) 
     \nonumber \\
   &&  \times \left[(F_1+F_2)\left(H_{Tg}(x,\xi)
     + {\Delta^2\over 4M^2}E_{Tg}(x,\xi)\right)n^\alpha
     + (F_1E_{Tg}(x, \xi) -F_2H_{Tg}(x,\xi)){\bar P^\alpha\over M^2} \right]
       \nonumber \\
    && + ~ \left(\Delta^\mu_\perp g^{\alpha\nu}_\perp
          + \Delta^\nu_\perp g^{\alpha\mu}_\perp 
          - g^{\mu\nu}_\perp\Delta^{\alpha}_\perp\right)
       { \alpha_s\over 4\pi} \left(\sum_q e_q^2\right) 
       \int^1_{-1} dx \alpha(x) \nonumber \\
   &&  \times \xi(F_1+F_2)\left(H_{Tg}(x,\xi)
     + {\Delta^2\over 4M^2}E_{Tg}(x,\xi)\right)
     \ . 
\end{eqnarray}
Here the presence of $\Delta$ in $\ell^{\mu\nu\alpha}$ and
$\Delta^\mu\Delta^\nu$ in $H^{(\mu\nu)\alpha}$ can give rise to a distinct
$\cos3\phi$ terms in the cross section \cite{diehl}.  To obtain the
latter, one just multiplies the lepton and hadron tensors together to get
the square of the $T$ matrix. Because of its length, we omit the final
expression. 

A more direct way to see the angular dependence of the cross section is to
use the formulas derived in the center-of-mass frame in Refs.
\cite{diehl,kro}.  According to these works, all one needs is the hadron
helicity amplitude $M^{\lambda,\lambda'}_{H,H'}$, where $\lambda$,
$\lambda'$ and $H$, $H'$ are the initial and final photon and nucleon
helicities, respectively.  The helicity-flip nucleon amplitude
$M^{-1,1}_{H,H'}$ clearly is just the helicity-flip Compton amplitude,
\begin{equation}
     M^{-1,1}_{h,h'}
    = \epsilon_\alpha(-1) \epsilon_\beta^*(+1)
      T^{\alpha\beta}_{\Delta \lambda = 2} \ . 
\end{equation}
This can readily be evaluated using Eq.(\ref{long}) by substituting in
the appropriate Dirac spinor for the nucleon helicity states. According to
\cite{diehl}, certain angular weighted cross sections can be used
to make a direct extraction of the above amplitude. 

\section{summary and comments}

In this paper, we have presented a number of new results related to the
helicity-flip off-forward parton distributions.  First, we enumerated
systematically all leading-twist off-forward parton distributions for a
nucleon: six for the quark parton and another six for the gluon.  Four of
these distributions, two each for the quark and gluon partons, involve
parton helicity-flip. Second, we derived the leading-logarithmic evolution
equations for these helicity-flip disitributions. In the forward limit,
our result agrees with the known kernel for $h_1(x)$ while the kernel for
$\Delta(x)$ is new. Third, we obtained the photon helicity-flip Compton
amplitude in terms of a tower of gluon operators with total derivatives.
Our result may be obtained from the known forward case by using the
conformal symmetry of QCD. Finally, we compute the leading DVCS cross
section which depends on the gluon helicity-flip distributions. 

We have emphasized the unique role played by helicity-flip distributions
in characterizing the properties of the nucleon.  If one askes for a clear
experimental signal of existence of gluons in the nucleon, the
helicity-flip Compton amplitude would serve the purpose. Without the
vector gluons, it would be at least power suppressed in the high-energy
limit.  Of course, the helicity-flip gluon distributions can also be
measured in vector meson production \cite{collins}. The size of the
helicity-flip distributions should be similar to the usual
helicity-dependent parton distributions -- there is no extra suppression
in the soft physics to curb helicity flip. 

\acknowledgements 
We would like thank J. Osborne for discussions and for drawing the Feynman
diagrams. PH thanks the Fulbright Foundation for sponsoring his visit to
the University of Maryland. This work is supported in part by funds
provided by the U.S.  Department of Energy (D.O.E.) under cooperative
agreement DOE-FG02-93ER-40762.


\begin{references}
\frenchspacing

\bibitem{ofpd}
K. Watanabe, Prog. Theo. Phys. 67, 1834 (1982); \\
F. M. Dittes et. al., Phys. Lett. B 209 (1988) 325; \\
D. M\"uller et. al., Fortschr. Phys. 42 (1994) 101.  

\bibitem{xdj}
X. Ji, Phys. Rev. Lett. 78, 610 (1997).

\bibitem{rad}
A. V. Radyushkin, Phys. Lett. B380, 417 (1996); Phys. Lett.
B385 (1996) 333; Phys. Rev. D56, 5524, 1997. 

\bibitem{xdj1}
X. Ji, Phys. Rev. D55, 7114 (1997).

\bibitem{hood}
P. Hoodbhoy, Phys. Rev. D56, 388, 1997. 

\bibitem{bal}
I. I. Balitsky and A. V. Radyushkin, Phys. Lett. B413, 114 (1997). 

\bibitem{blu}
J. Bl\"umlein, B. Geyer, and D. Robaschik,
Phys. Lett. B406, 161 (1997). 

\bibitem{mel}
X. Ji, W. Melnitchouk, and X. Song, Phys. Rev. D56, 5511 (1997). 

\bibitem{fran}
L. L. Frankfurt, A. Freund, V. Guzey, and M. Strikman,
hep-ph/9703449. 

\bibitem{pet}
V. Yu. Petrov et al., hep-ph/9710270.

\bibitem{bel0}
A. V. Belitsky and D. M\"uller, hep-ph9709379. 

\bibitem{bel}
A. V. Belitsky, B. Geyer, D. M\"uller, and A. Sch\"afer, hep-ph/9710427.

\bibitem{man}
L. Mankiewicz, G. Piller, E. Stein, 
M. Vanttinen, T. Weigl, hep-ph/9712251.
                                                                
\bibitem{ralston}
J. Ralston and D. Soper, Nucl. Phys. B152, 109 (1979). 

\bibitem{bel}
A. V. Belitsky and D. Mu\"ller, hep/ph9709379. 

\bibitem{art}
X. Artru and M. Mekhfi, Z. Phys. C45, 669 (1990). 
 
\bibitem{jaf}
R. L. Jaffe and X. Ji, Phys. Rev. Lett. 67 (1991) 552.

\bibitem{manohar}
R.L.Jaffe and A.V.Manohar, Phys. Lett. B223, 218 (1989).

\bibitem{collins}
J. C. Collins, L. Frankfurt, and M. Strikman,
Phys. Rev. D56, 2982 (1997).

\bibitem{man1}
L. Mankiewicz, G. Piller, and T. Wiegl, hep-ph/9711227. 

\bibitem{diehl}
M. Diehl, T. Gousset, B. Pire, and J. P. Ralston, 
Phys. Lett. B411, 193 (1997).

\bibitem{jaffe}
R. L. Jaffe, Nucl. Phys. B229 (1983) 205. 

\bibitem{diehl1}
M. Diehl and T. Gousset, DAPNIA-SPHN-98-01,  
e-Print Archive: hep-ph/9801233. 

\bibitem{bro}
S. J. Brodsky et al., Phys. Rev. D50, 3134 (1994). 

\bibitem{nzar}
M. Nzar and P. Hoodbhoy, Phys.Rev. D45, 2264 (1992).

\bibitem{gatto}
S. Ferrara, R. Gatto, and A. F. Grillo, Nucl. Phys. B34, 349 (1971). \\
Th. Ohrndorf, Nucl. Phys. B198, 26 (1982).

\bibitem{ap}
G. Altarelli and G. Parisi, Nucl. Phys. B126, 298 (1977). 

\bibitem{mul}
D. M\"uller, hep-ph/9704406. 

\bibitem{osborne}
X. Ji and J. Osborne, UMD PP\#98-074, hep-ph/9801260. 

\bibitem{kro}
P. Kroll, M. Schurmann, and P. A. M. Guichon, Nucl. Phys. 
A598, 435 (1996). 


\nonfrenchspacing
\end{references}
\end{document}